\documentclass[aps,prl,twocolumn,10pt, showpacs, longbibliography]{revtex4-1}
\usepackage{graphicx}
\usepackage{dcolumn}
\usepackage{bm}
\usepackage{graphicx}
\usepackage{color}
\usepackage{amsmath, amsthm, amssymb}
\usepackage[usenames,dvipsnames]{xcolor}
\usepackage{ulem}

\begin{document}
\title{Ultra-sensitive and Wide Bandwidth Thermal Measurements of Graphene at Low Temperatures}
\author{K.C. Fong and K.C. Schwab}
\affiliation{Applied Physics, California Institute of Technology, MC 128-95, Pasadena, California 91125}
\date{\today}
\begin{abstract}
{At low temperatures, the electron gas of graphene is expected to show both very weak coupling to thermal baths and rapid thermalization, properites which are desireable for use as a sensitive bolometer. We demonstrate an ultra sensitive, wide bandwidth measurement scheme based on Johnson noise to probe the thermal transport and thermodynamic properties of the electron gas of graphene, with a resolution of 2 mK/$\sqrt{Hz}$ and a bandwidth of 80 MHz.  We have measured the electron-phonon coupling directly through energy transport, from 2-30 K and at a charge density of $2\cdot 10^{11} cm^{-2}$.  We demonstrate bolometric mixing, and utilize this effect to sense temperature oscillations with period of 430 ps and have determined the heat capacity of the electron gas to be $2\cdot 10^{-21} J/(K\cdot \mu m^2)$ at 5 K which is consistent with that of a two dimensional, Dirac electron gas.  These measurements suggest that graphene-based devices together with wide bandwidth noise thermometry can generate substantial advances in the areas of ultra-sensitive bolometry, calorimetry, microwave and terahertz photo-detection, and bolometric mixing for applications in fields such as observational astronomy and quantum information and measurement.}
\end{abstract}
\maketitle

\section{Introduction}
Graphene is a material with remarkable electronic properties\cite{CastroNeto:2009p1599} and exceptional thermal transport properties near room temperature, which have been well examined and understood\cite{Balandin:2011p2120}. In fact, recent experiments have shown that graphene exhibits one of the highest phononic thermal conductivities of all measured materials\cite{Balandin:2008p1732, *Seol:2010p1987, *Nika:2009p2210}.  However at low temperatures, the thermodynamic and thermal transport properties are much less well explored\cite{Zuev:2009p1831, *Wei:2009p2105} and somewhat surprisingly, due to the single atomic thickness, low electron density, linear band structure, and weak electron-phonon coupling, the electron gas of graphene is expected to exhibit extreme thermal isolation \cite{Tse:2009p1892,Kubakaddi:2009p1895,Viljas:2010p1976}.

The very weak thermal coupling combined with exceptionally small electronic heat capacity of graphene leads to projections for very high sensitivity as both a bolometer\cite{Mather:1982p2060, Wei:2008p2106} and as a calorimeter \cite{MOSELEY:1984p2128, Roukes:1999p2061}. As is typical with extremely sensitive sensors, device readout with both the sensitivity and sufficiently low measurement back-action to realize the ultimate measurement sensitivity can also be a significant challenge.  Due to the thermal sensitivity and the relatively weak dependence of resistance on temperature\cite{Bolotin:2008p1770, Chen:2008p2054}, the simplistic use of electrical transport as the read-out scheme cannot realize the ultimate sensitivity of graphene given by thermodynamic fluctuations\cite{Mather:1984p2135,  Enss:2005p2167}. Here we present a measurement scheme based upon high frequency Johnson noise which provides both wide bandwidth and sensitivity to resolve the fundamental thermal fluctuations, with minimal disturbance to the thermal properties of the sample. This technique should be useful for both thermodynamic studies of graphene\cite{Gusynin:2005p2239, *Dora:2007p2241, *Trushin:2007p2228, *Long:2011p2240, Saito:2007p1844, *Lofwander:2007p2242, *Stauber:2007p2052, *Muller:2008p2222, *Foster:2009p2216, Ramezanali:2012p2232, Vafek:2007p2071} and for very low temperature bolometric applications.

This paper is organized as follows.  We first present the thermal model of the electron gas of graphene at low temperatures.  We discuss the Johnson noise measurement scheme and fundamental limits to the sensitivity.  We then present our measurements which provide a direct and accurate measurement of the electron-phonon coupling of graphene from 2-30K.  We present a measurement of the thermal time constant and determine the heat capacity of graphene.  Finally, we explore the expected sensitivity of graphene as a bolometer and as a calorimeter in the temperature range from 10~mK to 10~K.  These estimates suggest a number of exciting possibilities: detection of single microwave frequency photons, photon number resolution, and spectroscopy of terrahertz photons\cite{Wei:2008p2106, Kenyon:2008p2107, Karasik:2005p2223, Karasik:2011p2104}.

\section{Thermal Model}

\begin{figure*}
\begin{center}
\includegraphics[width=1\linewidth]{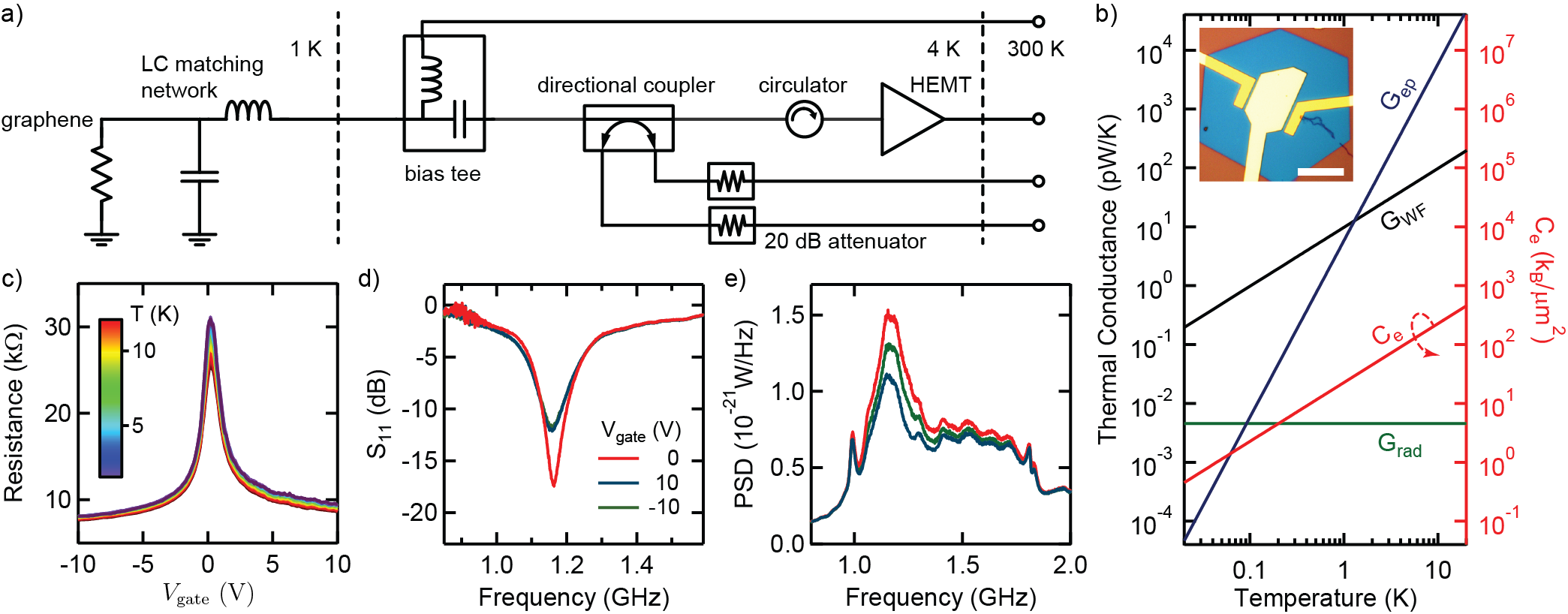}
\caption{\textbf{A} shows the measurement circuit: graphene, impedance matched with 1.161 GHz resonant, lithographic LC network (NbTiN film, $T_c=13.5$~K), and connected to HEMT amplifier.  \textbf{B} shows the expected thermal conductances ($G_{\rm{WF}}$, $G_{\rm{ep}}$,  $G_{\rm{rad}}$) and heat capacity versus temperature for  assuming a bandwidth of 80 MHz, $n=10^{11}$~cm$^{-2}$, and $A=10^{-10}$~m$^2$.  The inset shows an optical micrograph of the graphene sample. Scale bar is 15 $\mu$m long. \textbf{C} shows the two-terminal resistance of the graphene vs gate voltage, taken from 1.65-12 K.  \textbf{D} shows the reflected microwave response versus gate voltage.  The absorption dip at 1.161 GHz shows that the graphene is well matched to 50~$\Omega$ in a 80 MHz band. \textbf{E} shows spectra of the measured noise power taken at various sample temperatures, demonstrating the Johnson noise signal of the graphene in the impedance matching band.}
\label{fig:Setup}
\end{center}
\end{figure*}

Figure 1b shows the temperature dependence of the expected thermal conductance channels for the two-dimensional electron gas:  coupling to the electrical leads through electron diffusion, $G_{\rm{WF}}$, coupling to the lattice, $G_{\rm{ep}}$, and coupling to the electro-magnetic environment, $G_{\rm{rad}}$, where $G_{\rm{tot}}$  the sum of all three mechanisms.  

Thermal transport through electron diffusion in graphene has not yet been measured and most theories focus on the clean, low density graphene\cite{Trushin:2007p2228, Muller:2008p2222, Foster:2009p2216}. For doped samples dominated by the disorder potential of the SiO$_2$ substrate\cite{Zhang:2009p1757} such as the sample in this report, we assume the simple Wiedemann-Franz relationship as a starting point. $G_{\rm{WF}}=12L_0 T_e /R$, where $L_0$ and $R$ are the Lorenz constant and the electrical resistance of the sheet, respectively. The pre-factor, 12, is the result of the temperature profile developed by uniform ohmic heating and the presense of the thermal boundary condition of the contacts in a two terminal device (see comment [14] in Ref.~\cite{PROBER:1993p2102} and the supporting material).

The thermal conductance through the emission and absorption of blackbody photons into the electro-magnetic environment formed by the electrical measurement system is limited by the quantum of thermal conductance\cite{Schwab:2000p1852}, $G_0=\pi k_B^2 T_e/(6\hbar)$ and the bandwidth of the connection to the environment, $B$: $G_{\rm{rad}}=G_0(2\pi\hbar B/k_B T_e)$, assuming $B<k_B T_e/(2\pi\hbar)$ which is the bandwidth of the black body radiation\cite{Pendry:1983p2109, Meschke:2006p2110, Schmidt:2004p2111}.  

The electron gas can thermalize through the emission and absorption of acoustic phonons\cite{Roukes:1985p1900}. For temperatures below the Bloch-Gr\"{u}neisen temperature\cite{Hwang:2008p1899, Chen:2008p2054, Efetov:2010p2065} ($T_{BG}=2c\pi^{1/2}\hbar v_F n^{1/2} / (v_F k_B)$ = 33 K for $n=10^{11}$ cm$^{-2}$), heat transport between electrons and phonons in a doped sample in the clean limit ($k_p l_e\gg 1$ where $k_p$ and $l_e$ are the phonon wave vector and electron mean free path respectively) is expected to follow $\dot Q = A \Sigma (T_e^4-T_p^4)$\cite{Tse:2009p1892,Bistritzer:2009p1891,Kubakaddi:2009p1895,Viljas:2010p1976}, where  $A$ is the sample area, $\Sigma=(\pi^{5/2}k_B^4D^2n^{1/2})/(15\rho \hbar^4 v_F^2c^3)$ is the coupling constant, $D$ is the deformation potential, $v_F$ = 10$^6$ m/s is the Fermi velocity, c = 20 km/s is the speed of sound, and $n$ is the charge carrier density.  When $|T_e - T_p|\ll T_p$, $G_{\rm{ep}}=4 \Sigma A T^3$.  The electron-phonon coupling has been inferred using the temperature dependence of electrical transport for $T>T_{BG}$ at a charge density of $n=10^{12}$~cm$^{-2}$\cite{Chen:2008p2054} and for both $T>T_{BG}$ and $T<T_{BG}$ with very high charge density, $n\geq 10^{13}$~cm$^{-2}$\cite{Efetov:2010p2065}.  These measurements are consistent with the expected form and magnitude of the coupling.  

Through careful engineering of the sample geometry and the coupling to the electrical environment, it is possible to force the graphene to thermalize primarily through the electron-phonon channel, minimizing $G_{\rm{tot}}$ (see supplementary information) and at very low temperatures, this thermal conductance is expected to be extraordinarily weak\cite{Viljas:2010p1976} (Fig. 1).  The thermal modeling of our graphene sample shows the thermal conductance is expected to be dominated by $G_{\rm{ep}}$ for temperatures larger than 1 K, and by $G_{WF}$ for $T<1K$, as shown in Fig. 1b.   Superconducting leads can be used to block transport through $G_{\rm{WF}}$ \cite{Schwab:2000p1852} although care must be taken to avoid processes such as multiple Andreev reflection and electron-electron scattering that has been found to contribute to heat transport when using superconducting Al leads to contact graphene\cite{Voutilainen:2011p2043}. 

Furthermore, for the same reasons listed above which result in weak thermal coupling, the heat capacity of the electron gas is also expected to be minute. For doped graphene, the heat capacity is expected to be $C_e=(2\pi^{3/2} k_B^2 n^{1/2} T_e)/(3\hbar v_F)$\cite{Ramezanali:2012p2232, Viljas:2010p1976}. For perfectly undoped graphene, the heat capacity is expected to follow a $T^2\ln{T}$ temperature dependence\cite{Vafek:2007p2071}; this is not the situation for a disorder sample on SiO$_2$ substrate. At 100 mK and with $n=10^{11}$ cm$^{-2}$, one expects $C_e=2.3\cdot k_B$ for a $1\mu m$ x $1\mu m$ flake.  This combined with the thermal conductance, one can estimate the  thermal time constant: $\tau = C_e/G_{\rm{tot}}$.  Assuming $G_{\rm{tot}}\approx G_{\rm{ep}}$, one expects the maximum thermal time constant to be $\tau=10ps$ at 10K, $1ns$ at 1K, and $10\mu s$ at 10mK.  Due to the linear bands of graphene, and the correspondingly high Fermi temperature, the heat capacity of graphene can be 50 times lower than that of a heterostructure 2DEG, assuming $n=10^9$~cm$^{-2}$.
\begin{figure*}
\begin{center}
\includegraphics[width=1\linewidth]{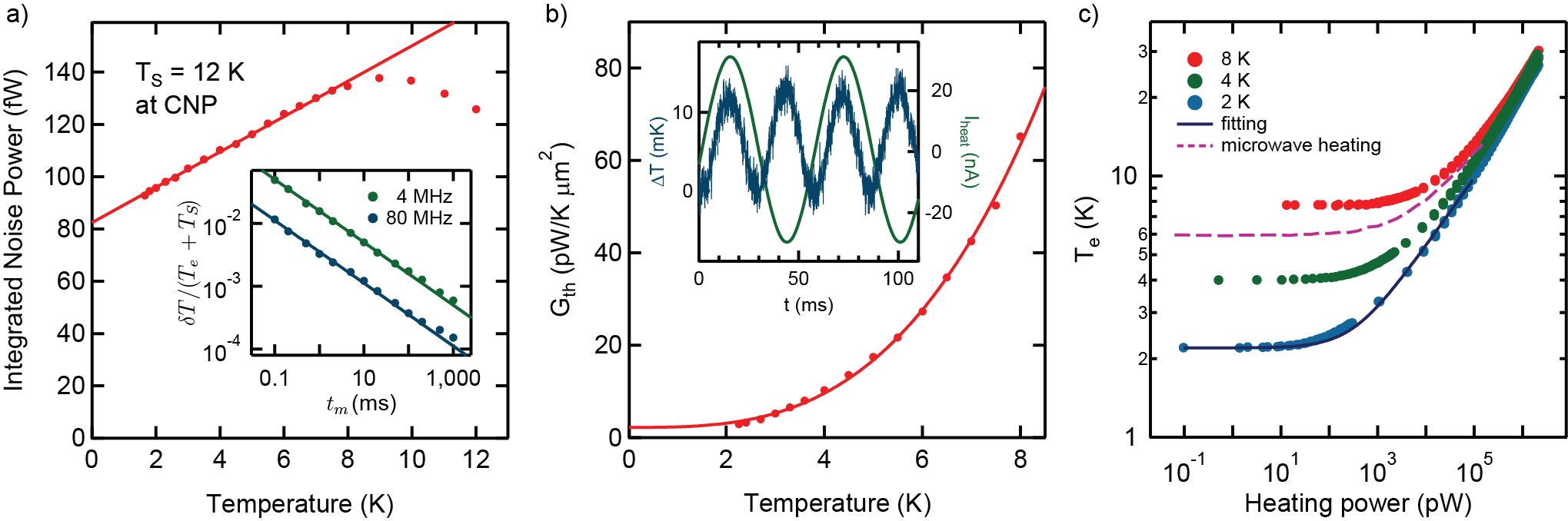}
\caption{\textbf{A} shows the integrated noise power vs refrigerator temperature, demonstrating the expected temperature dependence of the Johnson noise signal. The deviation at temperatures above 8K due to the temperature dependence of the NbTiN inductor.  The inset shows the precision of the noise thermometry taken with two measurement bandwidths, 4 and 80 MHz, vs integration time.  A resolution of 100 ppm is achieved, in agreement with the Dicke Radiometer formula (shown as lines).  \textbf{B} shows the results of the differential thermal conductance measurements.  The inset show a time trace, taken at 4 K, of the small heating current at 17.6 Hz, and the resulting 12 mK temperature oscillations at 35.2 Hz detected with the noise thermometer.  The red curve is a power law fit with exponent $2.7\pm0.3$. \textbf{C} shows the results of applying large dc heating currents at various sample temperatures, $T_p$ (points.) The expected form is shown as the blue line.  Also shown, is the heating of the electron gas vs applied microwave power at 1.161 GHz, also showing a similar heating curve and demonstrating the microwave bolometric effect with graphene.}
\label{fig:ThermalConductance}
\end{center}
\end{figure*}

\section{Noise Thermometry and Sample Fabrication}
Given the expected very weak coupling and high speed thermal response, we have implemented microwave frequency noise thermometry to explore these delicate and high bandwidth thermal properties, (Fig.1).  Noise thermometry has shown itself to be an excellent and nearly non-invasive probe of electron temperature for nanoscale devices with very minimal back-action heating\cite{Roukes:1999p2061, Schwab:2000p1852, Spietz:2003p2080}.  Curiously, the dominant effect of this measurement scheme on the graphene sample is to provide a thermal conductance channel for cooling through emission of photons into the measurement channel, $G_{rad}$.  This radiative channel will be present for any electrical readout scheme, however in bolometers which are sensed using electrical transport, heating due to ohmic loss is the dominant perturbation for an optimized measurement\cite{Mather:1982p2060, MOSELEY:1984p2128}.

The analysis of hot electron bolometers with resistive readout and operated with electrothermal feedback shows the optimized energy resolution to be $\Delta E\sim (2/\sqrt{\alpha})\cdot \Delta E_{th}$\cite{MOSELEY:1984p2128, IRWIN:1995p2134, Enss:2005p2167, Karasik:2011p2104}, where $\alpha=(T/R)dR/dT$ and $\Delta E_{th}=\sqrt{4C_ek_BT}$ is the energy resultion limited by thermodynamic fluctuations of the bolometer. At 2 K, we measure $\alpha \simeq 0.03$, which will limit the sensitivity of a resistively read-out graphene bolometer  to a sensitivity $12\cdot \Delta E_{th}$.  This situation becomes much worse at lower temperatures with both $T$ and $dR/dT$ decrease. Our analysis shows that an approach based upon noise thermometry is capable of approaching the thermodynamic limit, which leads us to believe that noise thermometry is preferable for experiments at very low temperatures.

Our graphene sample is fabricated using exfoliation onto a Si wafer coated with 285 nm of SiO; the single atomic layer thickness is confirmed using Raman spectroscopy\cite{Ferrari:2006p1830}. The high resistivity wafer (1-10 $\Omega$-cm at 300K) is insulating at cryogenic temperature to minimize the stray capacitance between the electrodes and ground, which would otherwise capactively load our impedance matching network. 

We match the relatively high impedance of a 15~$\mu m$~$\times$~6.8~$\mu m$ flake of graphene, $30~k\Omega$ at the charge neutrality point (CNP), to a 50~$\Omega$ measurement circuit using a lithographic, superconducting NbTiN LC network, which is placed a few millimeters from the graphene sample.  The LC network (L $\simeq$ 125 nH, C $\simeq$ 115 aF)  resonates at 1.161 GHz with a bandwidth of 80 MHz. As shown in Fig.~1d, absorption of microwave power is as high as 97\% within the matched bandwidth.  

In the same frequency band where we have engineered high absorption, the graphene is able to efficiently radiate Johnson noise, shown as the peak in the noise spectrums at 1.2GHz in Fig. 1e.  Analyzing these thermal noise spectrums below 8K in the matched bandwidth shows the expected linear dependence of the Johnson noise with temperature (Fig.~2a).  For $T> 8K$, the LC matching network begins to substantially shift its frequency as the temperature approaches the superconducting transition temperature and kinetic inductance effects become significant\cite{MESERVEY:1969p2198, Annunziata:2010p2199}. The temperature-axis intercept of Fig. 2a determines our system noise temperature of $T_S \simeq 12$ K. 

The sensitivity of our noise thermometry, $\delta T_e$, follows the Dicke Radiometer formula\cite{Dicke:1946p2064}: $\delta T_e / (T_e+T_S) = (Bt_m)^{-1/2}$, where $t_m$ is the measurement time, and $B$ is the measurement bandwidth.  For B = 80~MHz, this leads to a noise power density for electron temperature of $\sqrt{S_{T_e}}=2mK/\sqrt{Hz}$ at a sample temperature of 2 K. Figure 2a inset shows the measured normalized standard deviation of the noise temperature versus measurement time, plotted for two measurement bandwidths (4 MHz and 80 MHz); the data follows the Dicke Radiometer formula. We have recently realized much lower system noise temperatures ($T_s<$1K) with the implementation of a high bandwidth, nearly quantum-limited SQUID amplifier\cite{Muck:2001p2143}.

Ohmic contact is made using evaporated Au/Ti leads.  However, an advantage of very high frequency measurement is that high transparency contacts are not required since capacitive coupling to a metal film deposited on the graphene can also be utilized. A 100nm thick, Au electrostatic gate is insulated from the graphene using a 100 nm thick, over-exposed electron beam resist (polymethyl methacrylate) as a dielectric\cite{Duan:2009p2195, Henriksen:2010p2196} (area in blue in Fig.~1b inset). Fig.~1c shows the measured resistance of the graphene device versus gate voltage with mobility of approximately 3500 cm$^{2}$/Vs at low temperature, corresponding to a mean free path of about 20 nm. The charge carrier density at the CNP is approximately $2\times 10^{11}$ cm$^{-2}$, which is estimated by the width of the resistance maximum\cite{Bolotin:2008p1770, Dean:2010p2055}.

\begin{figure}
	\begin{center}
\includegraphics[width=1\linewidth]{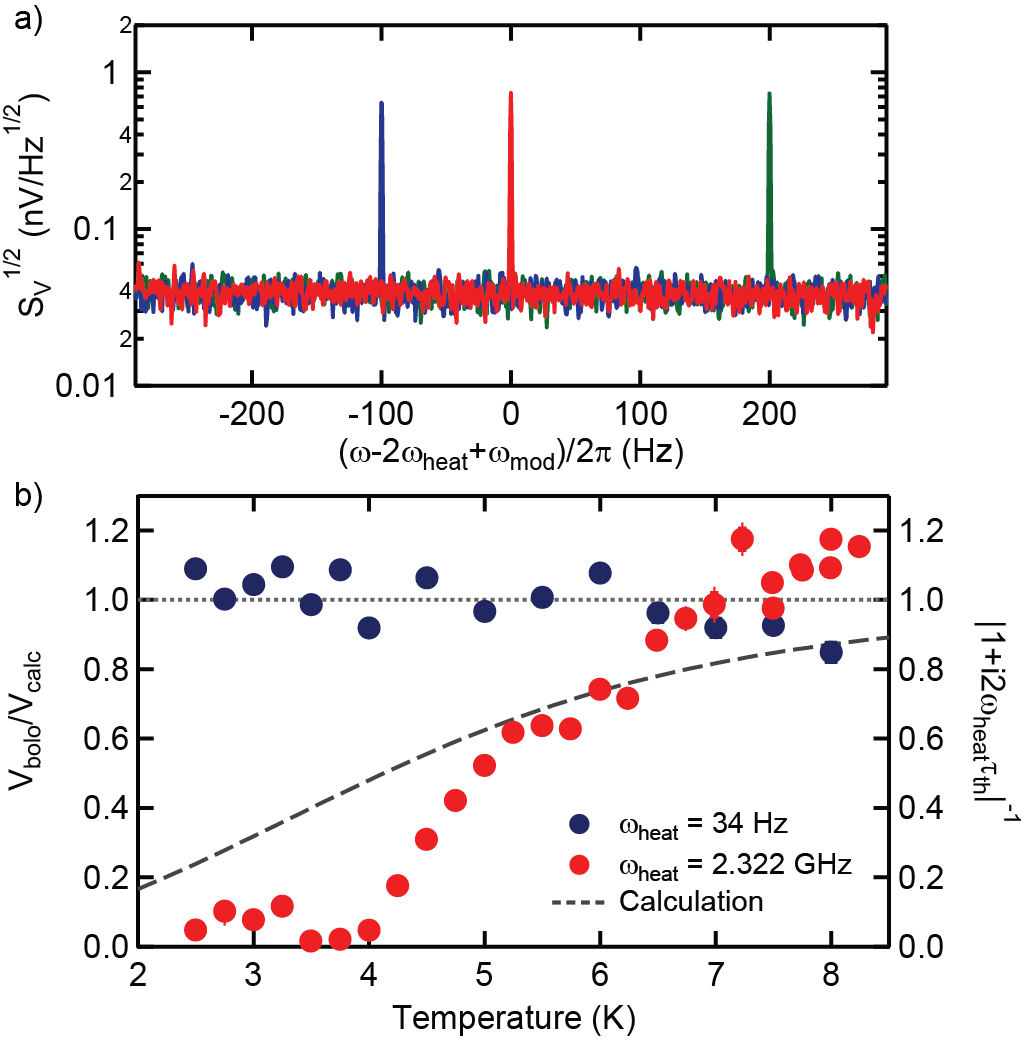}
\caption{\textbf{A} shows the spectrum, refered to the input of the HEMT, due to bolometric mixing.  The central, 600~pV red tone is due to a heating signal at $\omega_{heat}/2\pi =1.161$~GHz, which produces a temperature response at 2.322 GHz, which is then mixed down to $\omega_{bolo}=\omega_{heat}+2\pi\cdot 1$~kHz  using a small modulation tone at $\omega_{mod}=\omega_{heat}-2\pi\cdot 1$~kHz.  The blue spike, shifted down by 100 Hz, is the result of increasing the modulation tone by 100 Hz.  The green spike, shifted up by 200 Hz, is due to increasing the heating tone by 100 Hz, validating the expected relationship: $\omega_{bolo}=2\omega_{heat}-\omega_{mod}$.  \textbf{B} shows a plot of the measured bolometric mixing tone normalized by our expected signal assuming the graphene thermal time constant $\tau=0$.  The red points are measured with a heating tone of $\omega_{heat}/2\pi=1.161$~GHz, and blue points with $\omega_{heat}/2\pi=17.6$~Hz, both with a modulation tone of 1.161~GHz-1~kHz.  The dashed line shows he expected rolloff of the bolometric signal when $2\omega_{heat}/2\pi=2.32$~GHz and $\tau=C_e/G_{\rm{ep}}$.}
\label{fig:HeatCapacity}
\end{center}
\end{figure}

\section{Measurements}
We measure the thermal conductance of the electron gas by simultaneously applying currents through the graphene to produce ohmic heating, $\dot Q$, while measuring the electronic noise temperature. We have investigated the heat transfer in several limits: (1) measuring the differential thermal conductance with small quasi-static $\dot Q$ at various temperatures, (2) using larger dc current bias which produce temperature changes comparable or larger than the starting sample temperature, (3) and with microwave frequency heating currents. All of these measurements show similar results and are consistent with the existing theory of the electron-phonon coupling\cite{Kubakaddi:2009p1895, Tse:2009p1892, Bistritzer:2009p1891, Viljas:2010p1976}. 

Firstly, we impose a small oscillating current bias at 17.6 Hz, detect the resulting 35.2 Hz temperature oscillations of the electron gas in the limit where $\Delta T_e/T_e\simeq 10^{-2}$, and then compute the differential thermal conductance: $G=\dot Q / \Delta T_e$ (Fig.~2b).  This data is well fit with the expected form: $G=(p+1)\Sigma A T^p$, with $\Sigma$ and $p$ as fitting parameters, with values $0.07~W/(m^2K^3)$ and $2.7\pm0.3$ respectively. The power law exponent is near the theoretical expectation of $p=3$. Figure 2c also shows the results of applying a wide range of DC current biases such that $T_e$ can be much larger than $T_p$. We find this data to fit the expected form: $T_e=[\dot Q/(A\cdot \Sigma)+T_p^{p+1}]^{1/(p+1)}$  using $\Sigma$ and $p$ found from the differential measurements.

We also apply a heating signal at 1.161 GHz, at the center frequency of our LC matching network where the microwave power absorption into the graphene is nearly complete (Fig. 1d), and measure the increase in the electron gas temperature with the Johnson noise.  This method also shows the same thermal conductance as we found with quasi-static heating and demonstrates graphene as a bolometer to microwave frequency radiation. Using the measured Johnson noise thermometry sensitivity and thermal conductance, the noise equivalent power (NEP) of our graphene bolometer in this work is about 0.4 pW/$\sqrt{Hz}$ at 2 K.   Below we show that substantial improvements should be possible at lower temperatures and electron densities.

To probe the thermal time constant, $\tau$, and reveal the heat capacity of the graphene, we utilize the microwave frequency impedance matching network together with the small temperature dependence of the electrical resistance of graphene (Fig. 1c). This is a modified 3-$\omega$ method and is essentially a bolometric mixer\cite{PROBER:1993p2102}.  

We first apply a high frequency oscillating current, within the impedance matching band  to heat the sample: $\dot Q = I_{heat}^2(t)\cdot R(T_e)=I_{heat}^2 R(T_e) (1+cos(2\omega_{heat}t))/2$ where $\omega_{heat}=2\pi\cdot 1.161$~GHz. Similar to the above thermal conductance measurements, the DC component of the temperature change is observed through the Johnson noise of the sample. If the thermal time constant of the graphene electron gas satisfies $2 \omega_{heat} \tau<1$, then the temperature of the sample will oscillates at $2\cdot \omega_{heat}=2\pi\cdot 2.322$~GHz. From our measurement of the $G_{\rm{ep}}$ and our expectation of the heat capacity , we expect $2 \omega_{heat} \tau=1$ for $T\simeq 5.7$~K.   Given the weak dependence of the sample resistance on temperature, $dR/dT\sim 400\Omega /K$ at 2 K, the impedance, $Z(\omega)$, will also oscillate at $2\omega_{heat}$ with $T_e$.  By applying a second smaller modulation tone across the drain-source of the graphene sample at $\omega_{mod}=\omega_{heat}-1$~kHz,  the impedance oscillations are then transduced into a very small voltage oscillation, typically 10-100 pV, and are mixed back into the range  of our matching network: $\delta V(t) = (I_{mod}e^{-i\omega_{mod}t})\cdot(\delta Z e^{-i2\omega_{heat}t})$, where a component of $\delta V(t)$ oscillates at $2\omega_{heat}-\omega_{mod}$. Fig.~3a shows the mixing tone depends on the input frequencies as expected. See supplementary information for more details of the mixing and signal processing.  

For temperatures above 5 K we observe this mixed tone, resulting from bolometric mixing, with its amplitude agreeing with our expectations, demonstrating the temperature oscillations of the graphene sheet at 2.322 GHz.  However, for temperatures below 5 K we observe a substantial decrease in the amplitude of the mixed tone, consistent with the expected roll-off due to the finite thermal r				esponse time of the sheet due to the heat capacity (Fig.~3b). As a control of the experiment, we performed the same measurement procedures with $\omega_{heat}$ set to 2$\pi\times$ 17.6 Hz, and $\omega_{mod}= 2\pi\cdot 1.160999$ GHz. No roll-off of the bolometric mixed tone at $\omega_{mod}+2\cdot2\pi\cdot17.6Hz$ versus temperature was observed which is as expected since $2\omega_{heat}\tau_{th} \ll 1$ in this case. At 5.25 K, we calculate a heat capacity of 12,000 $k_B$ which is comparable to the smallest heat capacity measured to date\cite{Wei:2008p2106}.  We also believe this to be the first measurement of a two dimensional electron gas at zero field.

\begin{figure}
\begin{center}
\includegraphics[width=1\linewidth]{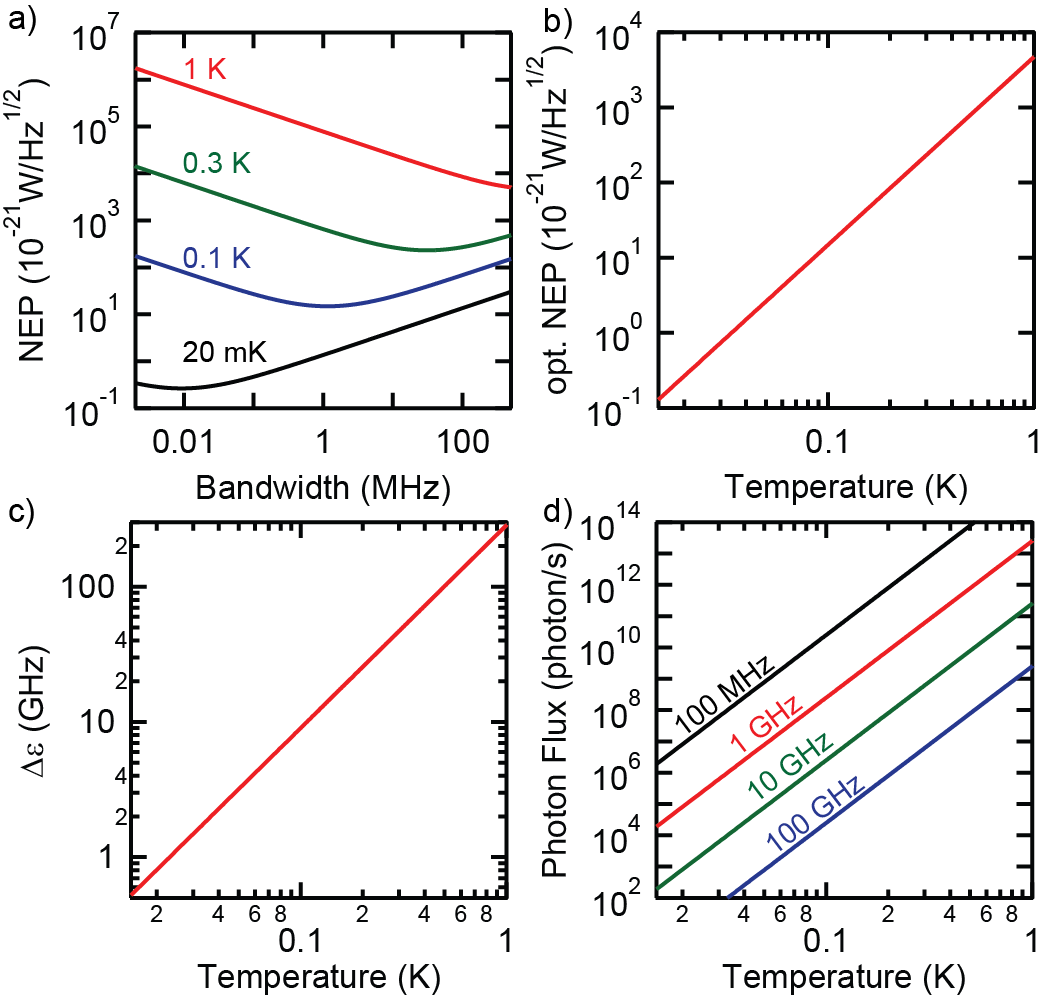}
\caption{\textbf{A} shows the expected sensitivity as a bolometer assuming $n=10^9$~cm$^2$ and  $A=10^{-11}$~m$^2$ versus coupling bandwidth, for various cryogenic operating temperatures, with \textbf{B} showing the optimal value versus temperature.  \textbf{C} shows the expected energy sensitivity to single photons. \textbf{D} shows the threshold for detection of shot noise of an incident microwave field of various frequencies.}
\label{fig:Sensitivity}
\end{center}
\end{figure}

\section{Discussion}
We compare our thermal conductance data to the existing theoretical predications of the electron phonon coupling \cite{Tse:2009p1892,Kubakaddi:2009p1895,Viljas:2010p1976}.  Using 30 eV as the deformation potential from the electrical transport measurements in single layer graphene\cite{Bolotin:2008p1770, Chen:2008p2054, Efetov:2010p2065}, we find our data consistent with an effective charge carrier density of $4.9\times 10^{11}$ cm$^{-2}$.  This is within a factor of 2.5 of the charge density estimated from resistance versus gate voltage measurements ($2\times 10^{11}$ cm$^{-2}$).  The agreement of our data to the theory, both in magnitude and temperature dependence, is somewhat surprising.  Firstly, we make our measurements at the CNP where the electron density is a result of a large disorder potential typical with SiO$_2$ substrates\cite{Zhang:2009p1757}.  Furthermore, our sample is deep in diffusive limit, i.e. $k_pl_e \ll 1$.  It is known from works in conventional two-dimensional electron gas structures that this diffusive limit and screening can substantially alter the electron-phonon coupling\cite{Sergeev:2000p2208}. To date, we aware of no published literature on the electron-phonon coupling in graphene in the diffusive limit and near the CNP. Theoretical work in this limit will be very useful.

The Wiedemann-Franz thermal conductance estimated here is based on that of a two-dimensional electron gas\cite{Gusynin:2005p2239, *Dora:2007p2241, *Trushin:2007p2228, *Long:2011p2240}. At low charge carrier density, theories have suggested deviations from the WF relationship due to the relativistic band structure\cite{Saito:2007p1844, *Lofwander:2007p2242, *Stauber:2007p2052, *Muller:2008p2222, *Foster:2009p2216}. This physics can be probed by performing this experiment at lower temperature.

The data we have gathered on the thermal transport and thermodynamic properties of graphene from 2-30 K and the sensitivity of our wide bandwidth thermometry system motivates an estimation of the sensitivity of graphene as a bolometer and photon detector at lower temperatures\cite{Mather:1982p2060, MOSELEY:1984p2128}.  Figure 4 shows the expected sensitivity as a bolometer versus noise bandwidth for various temperatures, where the noise equivalent power (NEP) is given by: $NEP = G_{\rm{tot}}\cdot\sqrt{S_T}$, where $S_T$ is the noise spectral density of noise thermometer.  Given the minute heat capacity for T $<$ 1K, the temperature resolution is expected to be limited by the thermodynamic fluctuations of the energy of the electron gas \cite{Mather:1982p2060, Chui:1992p2113}: $\langle \Delta T_e^2\rangle=k_BT_e^2/C$, which gives $S_{T}(\omega) = 4\tau k_B T_e^2 /(C(1+ (\tau \omega)^2))$ .   The maximum sensitivity versus measurement bandwidth is result of the balance between gaining resolution in the noise thermometry by increasing the measurement band, and increasing the thermal response by decreasing $G_{\rm{rad}}$.  As is clear from Fig.~4a and 4b, a graphene-based bolometer may exceed the sensitivity of the current state of the art bolometers developed for far-infrared/submillimeter wave astronomy with a sensitivity of $6\cdot 10^{-20} W/\sqrt{Hz}$ and a thermal time constant of $\tau=300$~ms \cite{Kenyon:2008p2107}, an improvement in bandwidth of $\sim$5 orders of magnitude.

As a photon detector and calorimeter, the expected energy resolution is given by\cite{MOSELEY:1984p2128, Roukes:1999p2061}: $\Delta E = NEP \cdot\sqrt{\tau}$.  Given the exceptionally fast thermal time constant, one expects single photon sensitivity to gigahertz photons (Fig.~4c).  For astrophysical applications in terahertz spectroscopy, one expects an energy resolution of one part in 1000 at 300 mK for an absorbed 1 THz photon.  This satisfies the instrument resolution requirements for future NASA missions (BLISS) at 3He-refrigerator temperatures\cite{Benford:2004p2108}. Compare to the recently proposed superconducting hot-electron photon counter for THz application at 300 mK\cite{Karasik:2005p2223}, the NEP of graphene-based bolometers is about 10 times less sensitive. However, the energy resolution for graphene is expected to be about 7 times better. In this way graphene bolometers are a possible solution for low photon flux photon counting in THz regime\cite{Karasik:2005p2223}. At 10mK, the intriguing possibility to observe single 800 MHz photons appears possible.

Furthermore, for high rates of photon flux, $\dot{n}$, the quantization of the field produces shot noise on the incoming power: $S_{shot}=2(\hbar \omega)^2 \dot n$ $W^2/Hz$.  For sufficiently high rates of microwave photons, this noise will dominate the temperature fluctuations of the sample (Fig.~4d).  At 100 mK, and with 10 GHz photons, for fluxes greater than $10^{6}$ photons/s the noise of the bolometer should be dominated by the shot noise of the microwave field.  In this way, graphene would act as a photodetector for microwaves: square law response, absorptive, and sensitive to the shot noise of the incoming field.  We know of no other microwave detector which has these characteristics and would open the door to novel quantum optics experiments with microwave photons\cite{Rocheleau:2010p1586}.

Note: During the writing of this work, we have become aware to three other experimental works which touch on some of these concepts \cite{Yan:2011p2062, *Vora:12p2211, *Betz:2012p2136}.

We acknowledge help with microfabricated LC resonators from M.~Shaw, and helpful conversations with P.~Kim, J.~Hone, D.~Prober, E.~Henriksen, J.~P.~Eisenstein, A.~Clerk, P.~Hung, E.~Wollman, A.~Weinstein, B.-I.~Wu, D.~Nandi, J.~Zmuidzinas, J.~Stern, W.~H.~Holmes, and P.~Echternach. This work has been supported by the the FCRP Center on Functional Engineering Nano Architectonics (FENA) and US NSF (DMR-0804567).  We are grateful to G. Rossman for the use of a Raman spectroscopy setup.  Device fabrication was performed at the Kavli Nanoscience Institute (Caltech) and at the Micro Device Laboratory (NASA/JPL).

%
\end{document}